\documentclass[letterpaper, reprint, superscriptaddress, amsmath, amssymb, aps, pre, nofootinbib]{revtex4-1} 

\usepackage{graphicx} 
\usepackage[normalem]{ulem}  
\usepackage{mathptmx} 
\usepackage{helvet}
\usepackage{amsmath,amssymb,amsfonts}
\usepackage{color} 
\usepackage{placeins}

\newcommand{\hordiff}[2]{\textrm{d}#1 / \textrm{d}#2} 

\begin{document}

\preprint{APS/123-QED}

\title{Modeling the origin of urban output scaling laws}

\author{V. Chuqiao Yang}
\email[E-mail me at: ]{vcy@u.northwestern.edu}
\affiliation{Department of Engineering Sciences and Applied Mathematics, Northwestern University, Evanston, IL, USA}
\affiliation{Santa Fe Institute, Santa Fe, NM, USA}
\author{Andrew V. Papachristos} 
\affiliation{Department of Sociology, Northwestern University, Evanston, IL, USA}
\affiliation{Institute for Policy Research, Northwestern University, Evanston, IL, USA}
\author{Daniel M.~Abrams}
\affiliation{Department of Engineering Sciences and Applied Mathematics, Northwestern University, Evanston, IL, USA}
\affiliation{Northwestern Institute for Complex Systems, Northwestern University, Evanston, IL, USA}
\affiliation{Department of Physics and Astronomy, Northwestern University, Evanston, IL, USA}

\begin{abstract}
Urban outputs often scale superlinearly with city population.  A difficulty in understanding the mechanism of this phenomenon is that different outputs differ considerably in their scaling behaviors. Here, we formulate a physics-based model for the origin of superlinear scaling in urban outputs by treating human interaction as a random process. Our model suggests that the increased likelihood of finding required collaborations in a larger population can explain this superlinear scaling, which our model predicts to be non-power-law. Moreover, the extent of superlinearity should be greater for activities that require more collaborators. We test this model using a novel dataset for seven crime types and find strong support.\end{abstract}

\maketitle

%
\section{Introduction}

\begin{figure}[!t]
	\centering
	\includegraphics[width=0.9\columnwidth]{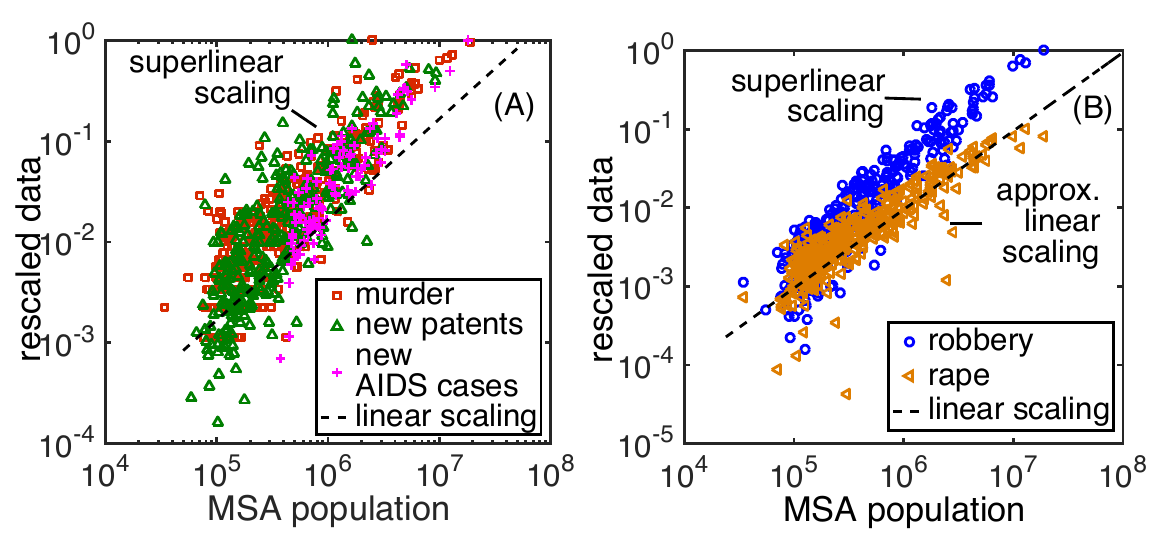} 
	\caption{\textbf{Urban outputs versus city size.} (A) Number of new patents, murder cases, and AIDS cases in U.S. Metropolitan Statistical Areas (MSA's) vs. population. These three urban outputs exhibit superlinear scaling---as MSA population doubles, the amount of output more than doubles. (B) Number of robbery cases and rape cases vs. MSA population. Robbery scales superlinearly, while rape scales close to linearly. All data are scaled to have maximum value 1.}
	\label{fig:data0}
\end{figure}

Physics-based models for human processes have had remarkable success in recent years, perhaps due to the increasing availability of relevant quantitative data (see, e.g., models of pedestrian synchrony, crowd dynamics, migration patterns, community formation, even changing religious affiliation \cite{strogatz2005, eckhardt2007, silverberg2013, karamouzas2014, lee2014, newman2004, abrams2011, mccartney2015}).  Here we explore the sociophysics of human productivity, focusing in particular on the origin of superlinear scaling laws that have been observed for a wide range of urban outputs (see Fig.\;\ref{fig:data0}A for several examples). Increases in these outputs can be mostly beneficial, as with GDP and patents, or mostly harmful, as with crime or contagious disease \cite{jacobs1961, bettencourt2007, bettencourt2016, rocha2015, hanley2016}.

\begin{figure}[!tbp]
	\centering
	\includegraphics[width=0.9\columnwidth]{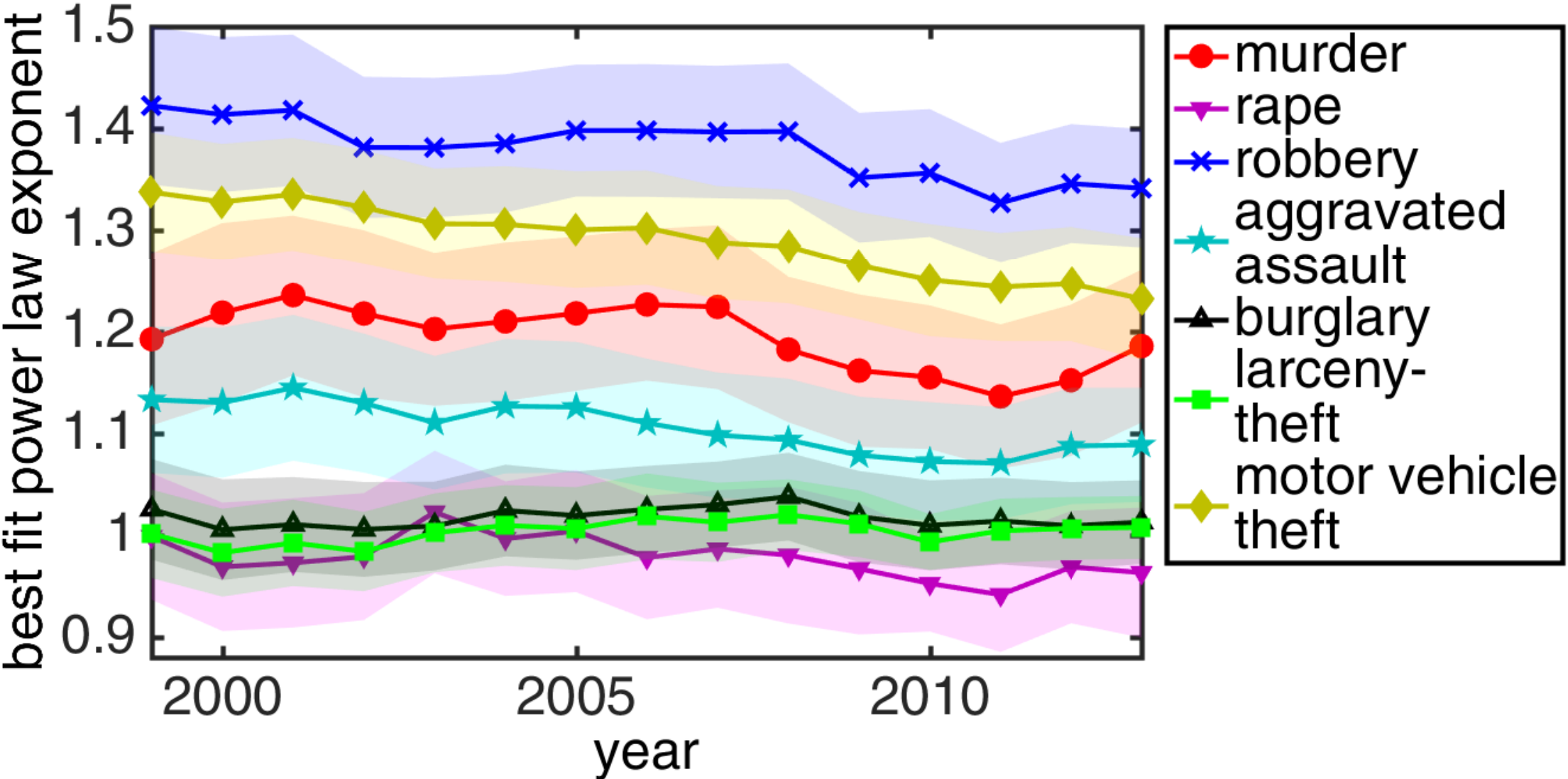}
	\caption{\textbf{Superlinear scaling over time.} Best-fit exponents (measuring degree of superlinear scaling) for scaling laws, i.e., slopes of curves such as those in Fig.\;\ref{fig:data0}, for all seven FBI crime categories. The shaded regions show 95\% confidence intervals. Exponents can differ considerably among crime categories, and are consistent over time.}
	\label{fig:all_year_fit}
\end{figure}

The scaling of serious crime was previously reported to be superlinear \cite{bettencourt2007, bettencourt2013} with a power law exponent of approximately $1.16$. When we break down the data and compare across the seven FBI crime report categories\footnote{FBI crime categories are murder, rape, robbery, aggravated assault, burglary, larceny-theft, and motor vehicle theft.}, however, the scaling behavior varies significantly: some categories show approximately linear scaling, while others are strongly superlinear. These differences persist for all years since 1999, the earliest year for which data are available (see Fig.\;\ref{fig:all_year_fit}). One illustrative example is the comparison between robbery and rape, as shown in Fig.\;\ref{fig:data0}B. Robbery scales superlinearly with city size, while rape scales close to linearly\footnote{Note that, for convenience, we frequently use the simpler but more ambiguous term ``city'' to refer to a Metropolitan Statistical Area (MSA) throughout this manuscript.}. 

It has remained unclear why some quantities are affected by city population more than others: previous efforts at understanding superlinear scaling in urban outputs have largely focused on the similarities rather than differences \cite{bettencourt2013, pan2013, arbesman2009, duranton2004, arbesman2011}.  In addition, many models \cite{ bettencourt2013, arbesman2009,duranton2004, arbesman2011, Gomez-Lievano2017} rely on a power law assumption for the scaling behaviors, which was recently challenged \cite{arcaute2015}. 

Explaining the variations in the scaling behavior and formulating a non-power-law framework are now two significant challenges in developing a scientific understanding of urban scaling. Here, we propose a novel model that explains and predicts the variation in scaling among different urban outputs, without relying on a power law hypothesis. 


\section{Mathematical model}
\subsection{Overview of the mathematical model}
Since most urban outputs, such as AIDS infection, patenting, and many types of crime have social components  \cite{cdc_hiv, lefevre1987, glaeser2011, warr2002, reiss1988, stolzenberg2008}, 
we are motivated to incorporate existing knowledge about social processes into our model. Mark Granovetter's landmark work ``The Strength of Weak Ties'' \cite{granovetter1973} argued that weak ties play an important role in providing information novel to one's social network that fosters outputs such as finding a job or starting a business. Motivated by this and direct empirical evidence for the importance of weak ties in innovation and crime \cite{hauser2007, patacchini2008}, we base our model on the assumption that finding the right collaboration is key to human productivity: one must meet all the necessary collaborators for an output in order to produce. Mathematically, this key concept is expressed as
\begin{equation} \label{eq:y}
  y(N) \sim N P_n[u(N)]\;, 
\end{equation}
where $y(N)$ denotes the volume of an urban output  (such as the total number of robbery cases or the number of patents) for a city with population $N$. The parameter $n$ is the number of partners needed for the output, $u(N)$ is the average number of unique contacts for a person living in the city, and $P_n[u(N)]$ is the probability of finding all required $n$ collaborators among $u(N)$ contacts. 

Here, we give an overview of the model's general conclusions, without functional form assumptions. In the next subsection, we will first show that $P_n$'s dependency on $u$ is in the form of
\begin{equation}\label{eq:p_n_relation}
P_n \sim u^n(N)\;. 
\end{equation}
Combining Eqs.~\eqref{eq:y} and \eqref{eq:p_n_relation}, we have, 
\[
y(N) \sim N u^n(N)\;.
\]
Taking the logarithm of both sides and differentiating with respect to $\ln(N)$, we have 
\begin{equation}\label{eq:dlogylogn}
\beta \equiv \frac {d \ln(y)}{d \ln(N)} \sim 1 + n \frac{d \ln(u)}{d \ln(N)}\;.
\end{equation}
Eq.~\eqref{eq:dlogylogn} leads to three predictions: 

\textbf{Superlinear scaling. }$\beta \equiv  {d \ln(y)}/{d \ln(N)}$ is often interpreted as the scaling exponent of $y$. Eq.~\eqref{eq:dlogylogn} predicts that if ${d \ln(u)}/{d \ln(N)} > 0$, meaning residents of more populated cities have more contacts (supported by empirical findings in \cite{schlapfer2014}), and $n>0$, meaning the activity typically requires more than one participant, then $\beta > 1$, giving rise to superlinear scaling.

\textbf{Variation in scaling exponents. }Eq.~\eqref{eq:dlogylogn} shows that $\beta$ increases with $n$. For fixed $N$, $\beta$ grows linearly with $n$. Thus Eq.~\eqref{eq:dlogylogn} predicts that urban outputs requiring more participants should exhibit more pronounced superlinear scaling. 

\textbf{Possibility of non-power-law superlinear scaling.} Individuals in bigger cities have the chance to meet more people, but cognitive limits (among other things) restrict them to only interacting with a small subset in a given period \cite{milgram1970}. It is thus plausible for ${d \ln(u)}/{d \ln(N)}$ to decrease for large $N$. Since the scaling exponent $\beta$ in Eq.~\eqref{eq:dlogylogn} may depend on $N$, the result is superlinear scaling behavior that is not a power law.

Considering typical patterns of collaboration can resolve the puzzle of why scaling behaviors vary across different urban outputs. The three predictions above hold for any general increasing function of $u(N)$. In the following sections, we first provide a derivation for $P_n(u) \sim u^n$. Then, in order to make quantitative predictions and compare the model with empirical data, we propose one general framework for estimating $u(N)$, derived from treating social interactions as a biased sampling process. 

\subsection*{Derivation for $P_n(u)$}
Among $u(N)$ unique contacts, only a small subset of those should result in partners for outputs such as crime and inventions. Whether a contact becomes a partner can depend on many factors: e.g., possession of a certain skill or establishment of a certain level of trust. We denote this probability by $\gamma$, which differs by the type of activity. 

The need to find all required collaborators for an output to occur can be interpreted in two ways. The first is that the output requires $n$ partners, each with a unique set of attributes (e.g., skill, relationship, etc.). The second is that the output requires $n$ individuals each possessing the same set of attributes. We present calculations for both interpretations and show that the scaling relationship for $P_n$, the probability of finding all partners needed, is $P_n \sim u^n$ (to leading order) for both interpretations. 


\textbf{Finding $n$ partners with unique attributes.}
The probability of finding at least one partner with a desired attribute out of the $u$ people met is
\[
 q(u, \gamma_i) = 1 - (1 - \gamma_i)^u\;,
\]
where $\gamma_i  \ll 1$ is the probability of any given individual having the attribute. 

After finding one partner among the $u$ individuals met (with probability $q(u, \gamma_1)$), then, if $n>1$, the searcher also needs to find another compatible partner among the $u-1$ remaining contacts (with probability $q(u-1, \gamma_2)$), and so on. The probability of finding all $n$ partners can be expressed as: 
\begin{equation} \label{eq:pn_1.1}
  P_{n} = q(u, \gamma_1) \cdot q(u -1, \gamma_2) \cdots q(u - n +1, \gamma_n) \;.
\end{equation}
We expand \eqref{eq:pn_1.1} assuming $\gamma_i \ll 1 $  for all  $i$. Since the expansion of $q(u, \gamma_i)$ near $\gamma_i = 0$ is $q(u, \gamma_i) \approx u \gamma_i + O(\gamma_i^2) $, the leading order term  for $P_n$ is:
\[
P_{n} \sim  u (u-1) \cdots (u - n+1)\prod_{i = 1}^n \gamma_i \;.
\]
With $u \gg n$, we have 
\[ 
P_{n}  \sim u^n \prod_{i = 1}^n \gamma_i \;.
\]

Because $\gamma_i$ are constants and we are only interested in how  $P_n$ scales with $u$, we express the scaling relationship as:
\[
 P_{n} \sim u^n\;. 
\]

\textbf{Finding $n$ partners with the same attributes.}
The probability of finding \textit{at least} $n$ compatible partners with the same attributes out of the $u$ people met is:
\begin{equation} \label{eq:pn_1}
	P_n (u) = \sum_{r= n}^{u}  \binom{u}{r} \gamma^r (1-\gamma)^{u-r} \;.
\end{equation}
where $\gamma$ is the probability of any person being a suitable partner. This calculation assumes the probability of each person being a suitable partner is independent.

The leading order term in Eq.~\eqref{eq:pn_1} is the first term ($r = n$):

\begin{equation} \label{eq:pn_2}
	P_n \sim   \binom{u}{n} \gamma^n (1-\gamma)^{u-n} 
	 =  \frac{u! \gamma^n (1-\gamma)^{u-n}} {n! (u-n)!} \;.
\end{equation}

We can simplify the ratio of factorials by writing it in terms of gamma functions:
\begin{equation}\label{eq:pn_1_1}
\frac{u!}{(u-n)!}  = \frac{\Gamma(u+1)}{\Gamma(u-n+1)}\;,
\end{equation}
which can then be approximated using Stirling's approximation \cite{tricomi1951}
\[
\frac{\Gamma(z+a)}{\Gamma(z+b)} = z^{a - b} \left[1 + \frac{(a + b)(a + b - 1)}{2z}\right]\;
\]
%
%
for large $z$ and bounded $a,~b$. Setting $z = u$, $a = 1$, and $b =  1-n$, Eq.~\eqref{eq:pn_2} can be simplified as
\begin{equation} \label{eq:pn_3}
	P_n \sim \frac{\gamma^n (1- \gamma)^{u - n} (2u - n^2 + n) u^{n-1}} {2n!} \;.
\end{equation}
Assuming $u \gg n$, and retaining only leading order terms (highest power in $u$ and lowest power in $\gamma$), we get
\[
	P_n \sim \gamma^n u^n \;.
\]
So the leading order scaling behavior of  $P_n(u)$ with $u$ is:
\[
	P_n(u) \sim u^n\;.
\]

\subsection*{Derivation for $u(N)$ }
In order to make quantitative predictions, we provide a framework to estimate the expression $u(N)$. Since an MSA is defined based on social and economic integration, we approximate an MSA of population $N$ as a closed system with respect to social interactions: all people in the city have some probability of interacting with one another. 

\begin{figure}[tbp]
	\centering
	\includegraphics[width=1\columnwidth]{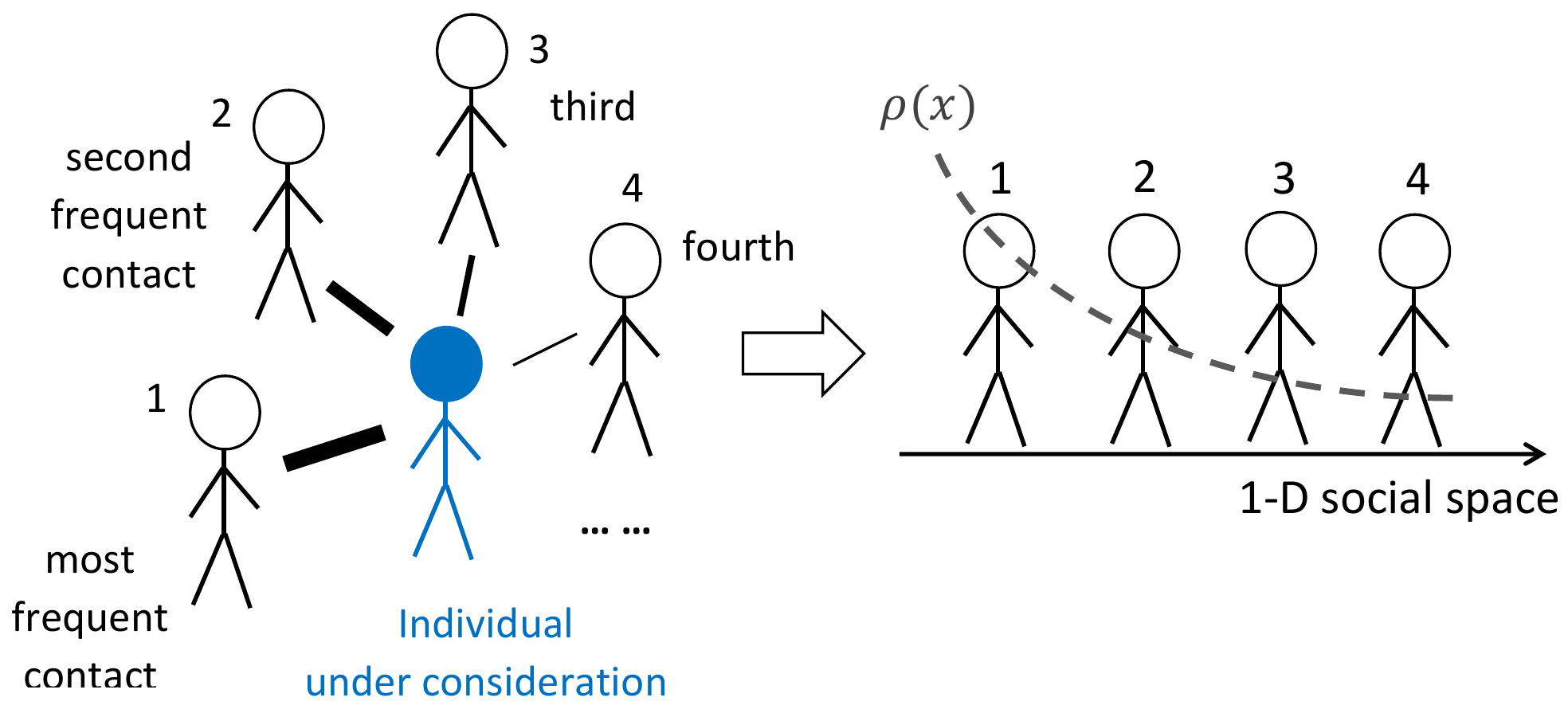}
	\caption{\textbf{Mapping to social space.} Cartoon for individuals in a city mapped to an one-dimensional space, ordered by rank of probability of social interaction.  The probability of interaction with a particular peer decreases with position in one-dimensional space $x$ as $\rho(x)$. }
	\label{fig:cartoon1}
\end{figure}

Spatial population distribution in cities is a complex problem on its own, and we wish to avoid assumptions about the spatial distribution of social interactions. Instead, we consider a ``social space"---a mathematical convenience to simplify our analysis---using the following approach. For each individual under consideration, we map all other individuals in the city to a one-dimensional space in which they are uniformly distributed and ordered by social distance to the individual under consideration (for an illustration, see Fig.\;\ref{fig:cartoon1}). A larger distance implies a smaller probability of interaction. The position of an individual in this social space corresponds to his or her rank based on social distance to (or probability of interaction with) the individual under consideration. We then treat each social interaction as a sampling process (independent and with replacement)---the person under consideration chooses a person in the social space with whom to interact with probability density function $\rho(x; N)$, where $x$ is position in the social space. By definition, $\rho$ is rank-probability distribution, and a non-increasing function of $x$.\footnote{In taking this ``social space'' approach, we leave for other work (see, e.g., \cite{simini2012} or \cite{grindrod2018}) the interesting questions of how geographical and social network structure may lead to particular scaling laws $\rho(x;N)$.}  

To simplify calculations, we first consider sampling segments of social space rather than individuals embedded in it. We discretise the one-dimensional social space of total length $L$ into $M$ patches of size $\Delta x$. The $i$th patch (with center position $x_i$) is thus chosen with probability $\rho(x_i) \Delta x$.  Taking $n_s$ to be the total number of samplings (interactions) that have occurred, the expected total unique space sampled, $L_u$ is   
\begin{equation} \label{eq:unique1}
  L_u  = \Delta x \sum^M_{i = 1} \left[ 
         1 - \left(1 - \rho(x_i)\Delta x \right)^{n_s} \right] \;.
\end{equation}

We denote by $L_s = n_s \Delta x$ the length of total space sampled with repeated samples counted cumulatively. We can then rewrite Eq.~\eqref{eq:unique1} as 
\begin{equation}\label{2}
  L_u  = \Delta x \sum_{i = 1}^{M} \left[1 - \left (1- \frac{\rho(x_i) L_s}{n_s}\right)^{n_s} \; \right] \;.
\end{equation}
The Laurent series expansion for $( 1 - c/n)^n$ as $n \rightarrow \infty$ is
\[
  \left(1 - \frac{c}{n}\right)^n = e^{-c} - \frac{c^2 e^{-c}}{2 n} + O\left(\frac{1}{n^2}\right)\;.
\]
Using this expansion, Eq.~\eqref{2} can be rewritten as 
\[
  L_u = \Delta x  \sum_{i = 1}^{M} \left[
    1 - e^{-\rho(x_i) L_s} + \frac{\rho(x_i)^2 L_s^2}{2n_s} e^{-\rho(x_i) L_s} + O\left(\frac{1}{n_s^2}\right)
  \right]\;.
\]
We take the continuum limit $\Delta x \rightarrow 0$ and $n_s \rightarrow \infty$, and neglect the $O(1/n_s)$ and higher order terms. Using $M \Delta x = L$, we have

\begin{eqnarray} \label{eq:Lu3}
  L_u &=& \lim_{\Delta x \rightarrow 0} \Delta x \sum_{i = 1}^M \left[1 - e^{-\rho(x_i) L_s}\right] \nonumber\\
      &=& M \Delta x - \lim_{\Delta x \rightarrow 0} \Delta x \sum_{i = 1}^M e^{-\rho(x_i) L_s}  \;.
\end{eqnarray} 
The second term of Eq.~\ref{eq:Lu3} is a Riemann sum. Taking the continuum limit of Eq.~\eqref{eq:Lu3} as $\Delta x \rightarrow 0$, the sum can be expressed in terms of an integral. We then have
\begin{equation}\label{eq:Lu}
  L_u = L - \int_0^L e^{-\rho(x) L_s} dx \;.
\end{equation}
Since the population distribution on the social space is uniform, the unique length covered by sampling ($L_u$) and the cumulative length sampled ($L_s$) directly correspond to the number of unique individuals met $u$ and the cumulative number of samples $s$ respectively. The total length of the rank-space $L$ by construction corresponds to the total population $N$. Changing notation in Eq.~\eqref{eq:Lu}, we find an expression for the number of unique individuals met $u$:
\begin{equation}\label{eq:u}
  u(N) = N - \int_1^{N} e^{-\rho(x) s} dx \;,
\end{equation}
where $s$ is the amount of sampling made by each individual in a certain period of time. Here we assume $s$ (reflecting casual contact interactions) does not change with city population\footnote{We note that little evidence exists for this hypothesis: much work has been done on friendship and acquaintance networks (e.g., \cite{milgram1970, hill2003} ), but little on scaling of casual contact or weak tie numbers.  We expect our hypothesis of constant $s$ to be conservative in the sense that, if there is some city size dependence, presumably $s(N)$ is an increasing function, which would result in even greater superlinearity than our model currently predicts.}. We will estimate its value (assumed universal for simplicity) by fitting to all available datasets.

\subsection{Closed-form expression for scaling of urban outputs}

The closed-form expression for the total output $y$ is thus 
\begin{equation} \label{eq:y2}
  y(N) \sim  N \left(N - \int_1^{N} e^{-\rho(x) s} dx \right)^n\;.
\end{equation}
The parameter $s$ is a measure of social capacity, or the amount of casual social interaction (including repeated interactions) in a characteristic time period. For simplicity we make the conservative approximation that $s$ is universal for all individuals (see earlier footnote). The function $\rho(x)$ is a rank-probability distribution representing the probability of interacting with an individual at rank $x$ in one's list of contacts sorted by contact frequency. Intuitively, $\rho(x)$ can be understood as representing a social interaction pattern: how often does one interact with one's most frequent contact vs.~one's second-most-frequent contact, third-most-frequent contact, etc. 
Importantly, it applies not just to close relationships, but also extends to casual contacts with whom one may not maintain any relationship; those are taken as seeds of new collaborations.

\subsection*{Secondary correction}

Other authors \cite{gould2002, grogger1997, machin2004} have argued that the incentive to commit crime drops with city size; we incorporate this effect as a secondary correction to our model: 
\begin{eqnarray} \label{eq:correction}
    y(N) & \sim & N \cdot u(N)^n N^{-0.12} \\ \nonumber
    & \sim & N^{0.88} \left( 
    N -  \int_1^{N}  e^{-\rho(x; N) s}  dx \right)^{n}~. \\ \nonumber
\end{eqnarray}
Note that this is simply a shifted version of the model in Eq.~\eqref{eq:y2}, and results reported below stay largely the same with either version (our theoretical predictions for power law fits are simply shifted globally by 0.12). See supplementary material section 7 for details on this secondary correction.

\section{Results and empirical evidence}
Even without assumption on the functional form of $\rho(x)$, Eq.~\eqref{eq:u} gives $d u / dN \geq 0$ \cite[see $\mathsection$8]{SMtext}. This implies that individuals with identical social capacities and social interaction patterns will (on average) meet more unique individuals in more populated cities. This result is consistent with empirical findings from phone contact networks in a number of cities \cite{schlapfer2014}.

In order to get quantitative predictions, we need to make an assumption about the form of $\rho(x;N)$. Motivated by observations of Zipf's law scaling in a variety of rank distributions (e.g., word frequency, city population, earthquake magnitudes \cite{newman2005}), and direct evidence supporting the hypothesis that communication networks (such as emails, phone calls and face-to-face interactions) have power-law like degree distribution \cite{ebel2002, aiello2000, cattuto2010}, we assume $\rho(x)$ to have the following form:
\begin{equation}   \label{eq:rho}
  \rho(x;N) = m(N) x^{-\alpha}~\;,
\end{equation}
where $m(N)$ is a normalization factor such that $m(N)\int_1^N x^{-\alpha} dx = 1$. We will fit the parameter $\alpha$ when validating with urban scaling data, and also use an independent dataset (the communication patterns in the Enron email corpus) to check that the parameter found is in a reasonable range. Note that we also consider other options for the algebraic form of $\rho(x)$ and find similar results (see \cite{SMtext}, $\mathsection$3).

\begin{figure}[!t]
\centering 
\vspace{-0.0in}
\includegraphics[width = 0.9\columnwidth]{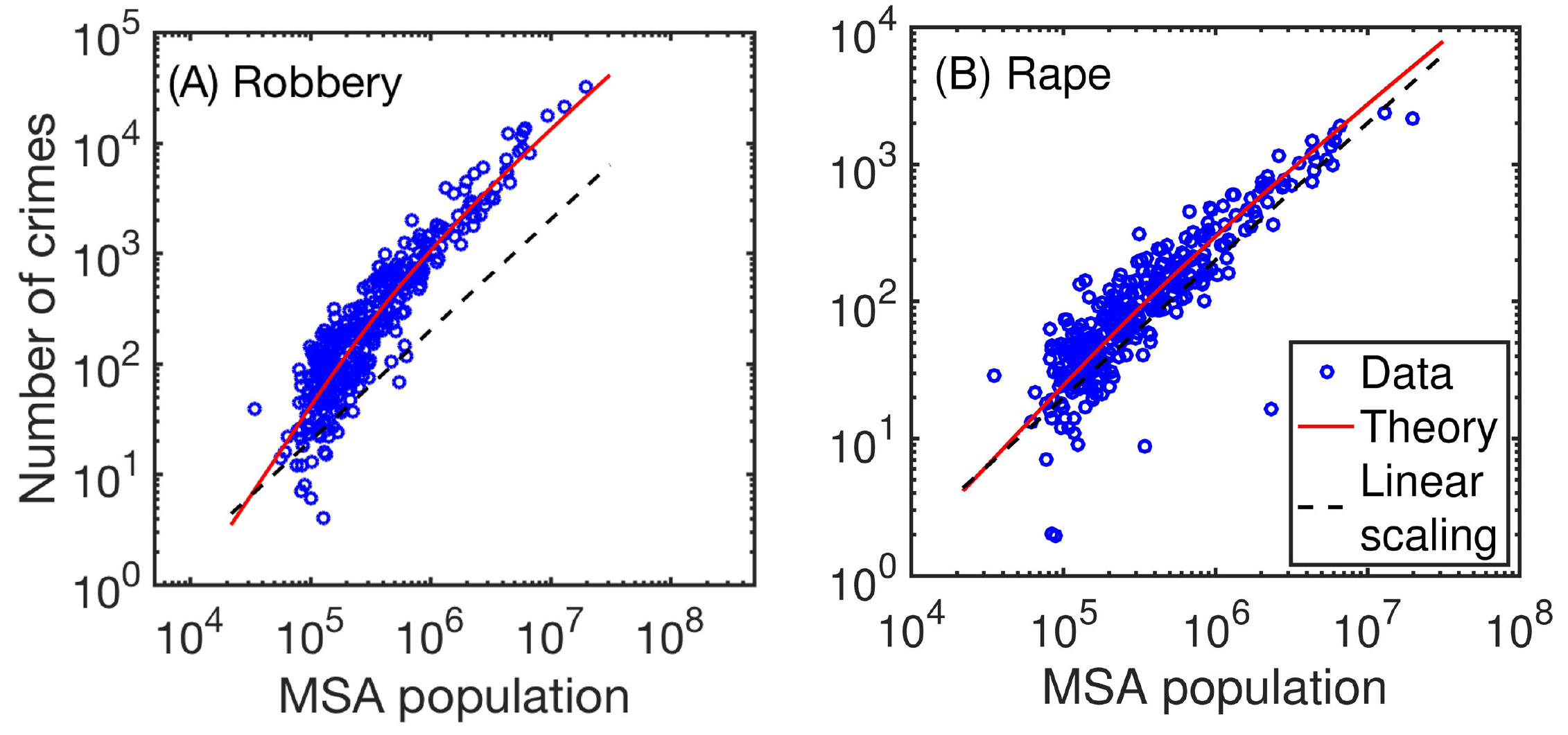}
\caption{\textbf{Model predictions with data.}  (A) Number of robbery cases versus MSA population. (B) Number of rapes cases versus MSA population.  Both for US in 2012. Red solid curves shows model predictions (not power law), blue dots show data points, and the black dashed curves are reference lines of unit slope. Our model predicts superlinear scaling for robbery, and approximately linear scaling for rape.  The difference between the model predictions in (A) and (B) is a result of the differing average co-offending group sizes. This figure uses co-offending group size calculated from the NIBRS dataset. The average group sizes are 1.74 for robbery and 1.29 for rape (group size is $n+1$). The parameters used in the model's prediction are $s = 2.63 \times 10^6$ and $\alpha = 0.93$ (global fit to all types of crimes). 
}
	\label{fig:fits}
\end{figure}

The integral in \eqref{eq:u}, after plugging in Eq.~\eqref{eq:rho} can be approximated as an incomplete gamma function  $(s m)^{1/\alpha} \alpha^{-1} \Gamma(-\alpha^{-1}, m N^{-\alpha}s)$
\cite[see $\mathsection$8]{SMtext}. Combining that with Eq.~\eqref{eq:correction}, we reach a closed-form estimate for the scaling behavior of social output in a city of population $N$: 
\begin{equation}\label{eq:y3}
    y(N)  \sim N^{0.88} \left[N - \frac{(s \; m)^{1/\alpha}}{\alpha} \Gamma \left(-\frac{1}{\alpha}, m \;N^{-\alpha}s \right) \right]^{n} \;,
\end{equation}
where $m  = m(N)= (\alpha - 1)/(1 - N^{1 - \alpha})$ if $\alpha \neq 1$; $ m = 1/\ln(N)$ if $\alpha = 1$. The parameter $n$ is the typical number of partners needed for an output. We input this parameter's value from data on average co-offending group size in the National Incident-Based Reporting System (NIBRS).


\subsection{Support from empirical data}
\begin{figure}[!t]
	\centering
	\vspace{0.0in}
    \includegraphics[width = 0.95\columnwidth]{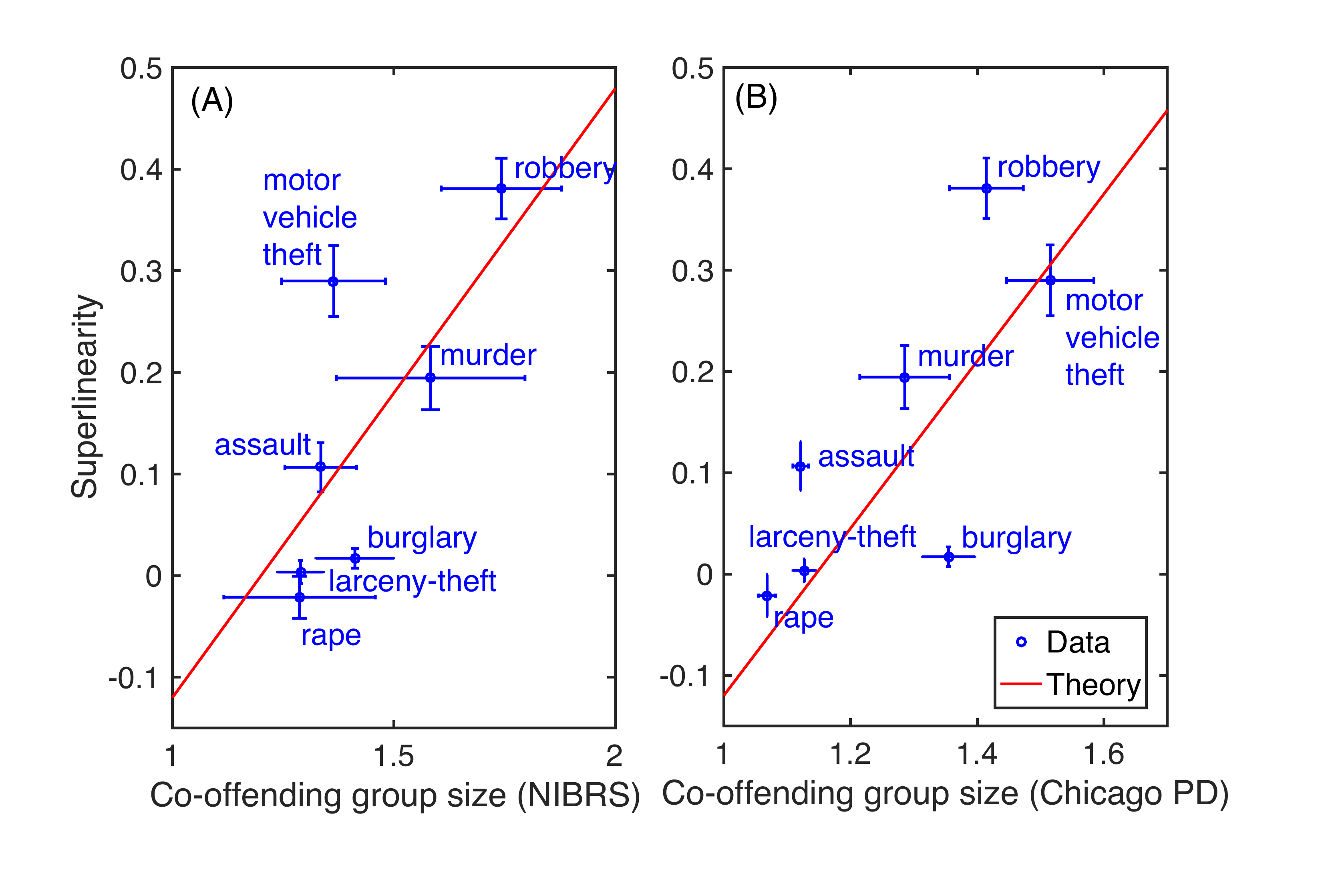}
	\caption{\textbf{Superlinearity as a function of average group size for crimes.}
	Superlinearity is quantified as the exponent of the best-fit power law minus 1; this is for convenience---our model does not predict power law scaling. Red solid line shows model prediction. Vertical error bars in both panels are standard deviation of year-to-year variation over 1999--2012. In (A), the horizontal error bars are state-to-state standard deviation in mean group sizes.  In (B), the horizontal error bars are standard deviation of year-to-year variation over 1999--2012. Sources: \cite{FBI,bettencourt2004, NIBRS}. }
	\label{fig:punch2}
\end{figure}

Our model agrees well with US FBI data on all seven crime categories across 14 years; typical comparisons are shown in Fig.~\ref{fig:fits} (see \cite[$\mathsection$5]{SMtext} for all comparisons) and a summary is shown in Fig.~\ref{fig:punch2}. The model explains not only the observed superlinear scaling for some urban outputs (e.g., robbery in Fig.\;\ref{fig:fits}-A), but also close-to-linear scaling in others (e.g., rape in Fig.\;\ref{fig:fits}-B).  In Fig.~\ref{fig:punch2} we show the relationship between the average co-offending group size and the degree of superlinearity (quantified for ease of comparison by the best fitting power-law exponent to the scaling relation minus one---though note that our model \textit{does not} predict power law scaling). The two panels in Fig.~\ref{fig:punch2} use two independent sources of data for the co-offending group size---NIBRS in  Fig.~\ref{fig:punch2}-A and Chicago Police Department in Fig.~\ref{fig:punch2}-B. The Pearson correlation between average co-offending group size and superlinearity in the data is 0.764 (p-value 0.046) in Fig.~\ref{fig:punch2}-A, and 0.761 (p-value 0.047) in Fig.~\ref{fig:punch2}-B. The Spearman's rank correlation is 0.821 (p-value 0.034)in Fig.~\ref{fig:punch2}-A and 0.786 (p-value 0.048) Fig.~\ref{fig:punch2}-B. 


\subsubsection{Background on empirical data}
Average co-offending group sizes for the seven types of crimes come from crime reports of two independent sources: the National Incident-Based Reporting System in 2014 and the Chicago Police Department arrest records in 1999-2012 (see \cite[$\mathsection$1]{SMtext} for more detail on those sources). Both sources report incident-level records for a variety of crime types. The co-offending group size of each incident is defined as the number of unique offenders reported in that incident. We average over all incidents of each type to reach the average group size. The parameter $n$, the number of partners is calculated by average group size minus 1.  Note that average group sizes vary only over a small range for our crime data.  This is primarily due to inherent limitations in offender reporting: for many crimes committed by groups, only a subset of the co-offenders are arrested or listed in crime reports (a single arrestee is the most common case).  Despite this limitation, the data do show significant and consistent variation in group sizes, but the average values we use should be interpreted as correlates of, rather than direct estimates of, true co-offending group sizes.\footnote{As long as data-derived co-offending group sizes are monotonically increasing functions of the true co-offending group sizes, we expect correlations between model predictions and data to be preserved.} To ensure the robustness of our results, we use both sources for model validation and find support of our prediction from both data sources.

\subsubsection{Parameter fitting, model selection, and robustness}

When comparing our model to data, we use a total of 98 crime scaling datasets (7 types $\times$ 14 years). Two parameters are assumed to be the same across all datasets, $\alpha$ and $s$, describing a social interaction pattern and ``social capacity,'' respectively. Each dataset has another proportionality constant that is fitted. We find the values of the two global parameters ($s$ and $\alpha$) by minimizing the sum of the 2-norm error across all 98 datasets. The parameter $n$ is input from the average co-offending group size data for each type of crime. The best fitting global parameter pair was $s = 2.6 \times 10^6$, $\alpha =  0.93$ for the NIBRS dataset, and $s = 4.0 \times 10^6$, $\alpha = 0.69$ for the Chicago dataset. Our model performs better than power-law models (as measured by AIC and BIC \cite[see $\mathsection$6]{SMtext}) and has fewer fitted parameters\footnote{We compare our model with the power law assumption ($y = a N^b$) for each data set. For $k$ data sets, the power law model requires fitting $2k$ parameters, while ours only requires $k + 2$ .}, suggesting that our framework may be valuable for understanding this and similar phenomena.

We note that our model predictions are generally quite robust to the choice of $\rho(x)$.  Any decreasing function $\rho(x)$ will imply $d u / dN \geq 0$, leading to the prediction of superlinearity. The predictions are not particularly sensitive to the details of the particular form of that function---in addition to power laws, we also tested truncated log normals and a piecewise-constant function (motivated by the ``circle of acquaintanceship" concept \cite{dunbar2008}), with nearly equivalent results (\cite[see $\mathsection$3]{SMtext} for details).  To avoid overfitting, we do not attempt to explore the space of all possible (or all plausible) functions $\rho(x)$, we choose what we see as the simplest, the power law.

\section{Discussion}
Some prior work has attributed superlinear scaling to the hypothesis of hierarchy in infrastructure and social networks \cite{bettencourt2013, arbesman2009}, or differences in population density \cite{pan2013}. Our model suggests that a simple ``finite-size effect,'' i.e., limited population to sample from in small to midsize cities, could be the key underlying mechanism. This may at first appear surprising, since even medium-sized cities in the U.S.~include hundreds of thousands of unique individuals.  At a plausible high rate of 100 ``sampling events'' per day, however, an individual would have nearly 1.5 million samples after 40 years, more than the population of all but the largest U.S.~cities (though of course the number of \textit{unique} individuals met will be far less).  

In our model, the finite-size effect reduces as city population becomes large:  Eq.~\eqref{eq:y} implies that $\hordiff{y}{N}$ decreases as $N$ increases, and as $N \rightarrow \infty$, $\hordiff{y}{N} \rightarrow 1$. Data such as that shown in Fig.\;\ref{fig:data0} display a reduced slope for the largest cities. This is consistent with the disappearance of this finite-size effect at the upper limits of U.S.~city size. This suggests that, with limited resources, populating smaller cities would have a bigger impact on overall urban productivity than populating already big ones. 

Some authors have taken alternative approaches to scaling of crime in cities, such as using a Bayesian framework \cite{gomez2012} or looking at empirical connections between crime and other urban indicators \cite{alves2014}. Others have discussed how urban outputs such as crime may display long-term memory \cite{bettencourt2010}, how temporal clustering relates to crime scaling \cite{hanley2016}, or how crimes cluster geographically in cities \cite{oliveira2017}. Our model is not mutually exclusive with these others. However, most efforts continue to operate under the assumption of power law scaling. We hope our work will encourage the study of crime outside of the power law framework. 

In \cite{clauset2009, virkar2014}, Clauset et al.~argue that commonly used statistical approaches to fitting and testing power laws can be problematic. Leitao et al.~\cite{leitao2016} recently examined a variety of power law models in this context, showing that data were often inconsistent with models, and that many estimated exponents were not statistically distinguishable from 1. Our model was partially motivated by the (perhaps) over-dependence on power law assumptions in the literature.  The scaling we predict could explain poor fits of data to power laws: inferred exponents would vary with the range of city size.

Depersin and Barthelemy \cite{depersin2018}, motivated by a longitudinal dataset on traffic congestion in cities, argue that scaling laws depend not only on population but also on growth history. Our model could also be generalized to incorporate such dependency---factors such as group size or social interaction patterns could show memory effects.

Our model, in agreement with previous models \cite{bettencourt2013, pan2013}, implies that the dual aspects of cities are not separable: both positive and negative urban outputs (e.g., inventions and crimes) share common driving mechanisms rooted in social interaction. Future research---especially in the study of crime, law, deviance, and other sources of urban inequality---would do well to consider how scaling models such as ours might be further calibrated to capture differences within cities, especially across neighborhoods or communities. 

This paper relies heavily on crime as an example because of the abundance and quality of data we were able to compile. However, the principle of the model can generalize to other urban outputs driven by forming collaborations, such as inventions and starting new businesses. We applied our model (with the same $\alpha$ and $s$ parameters found by fitting to the crime data) to patent scaling laws, while extracting empirical average group sizes from patent co-authorship. We find good agreement between our model and patent scaling behavior (see \cite[see $\mathsection$9]{SMtext} for details). We did not include the patent group size in the comparison in Fig.~\ref{fig:punch2} because the rate of under-reporting of group sizes likely differs between patents and crime. 

\section{Conclusions}
The good agreement between our simple model and data indicates that differences in scaling relationships can indeed result from differences in the typical number of participants for an urban output: those outputs that are more ``social'' in nature are more strongly affected by city population. In agreement with previous models \cite{bettencourt2013, pan2013, Gomez-Lievano2017}, we find that a fundamental driving mechanism of scaling in urban productivity is social interaction.

\begin{acknowledgments}
The authors would like to thank the Chicago police department for making data available, and Sara Bastomski and Jennifer Wu for help with the co-offending dataset. This research was partially supported by the James S. McDonnell Foundation through Award No. 220020230.

\end{acknowledgments}

\end{document}